# Reply to comment by K.A. Duderstadt et al. on "Atmospheric ionization by high-fluence, hard spectrum solar proton events and their probable appearance in the ice core archive"

Claude M. Laird[1], Adrian L. Melott[2], Brian C. Thomas[3], Ben Neuenswander[4], and Dimitra Atri[5],

[1] Retired, Lawrence, Kansas 66047 USA   email claude.m.laird@gmail.com

[2] Department of Physics and Astronomy, University of Kansas, Lawrence, Kansas 66045 USA email melott@ku.edu

[3] Department of Physics and Astronomy, Washburn University, Topeka, Kansas 66621 USA email brian.thomas@washburn.edu

[4] Specialized Chemistry Center, University of Kansas, Lawrence, Kansas 66047 USA email bennder@ku.edu

[5] Blue Marble Space Institute of Science, Seattle, Washington 98109 USA email dimitra@bmsis.org

Corresponding author:  B. C. Thomas (brian.thomas@washburn.edu)

**Key Points:**

1. Alleged discrepancies between predicted and measured nitrate deposition are grossly overestimated.

2. Simulating a different event does not demonstrate agreement with balloon measurements.

3. Signatures in nitrate and $^{10}$Be ice core records support the value of nitrate as an indicator of strong, hard-spectrum SPEs.

**Abstract**

*Duderstadt et al.* [2016b] comment that the *Melott et al.* [2016] study of nitrate formation by solar proton events (SPEs) and comparison with the ice core archive is "fundamentally flawed," because it does not include pre-existing $HNO_3$ in the stratosphere. We show that they exaggerate both the enhancement predicted by our findings and pre-industrial $HNO_3$ levels in their model, and fail to prove this assertion. Our feasibility study matched expected SPE nitrate production with ground truth measurements. It is not clear that their approach is more realistic and absence of a detailed mechanism does not disprove our results. Models can be no better than the information they are provided and in this case there continue to be significant unknowns and uncertainties, especially in the role of polar stratospheric clouds (PSCs) and possible interactions with cosmic rays that constitute lower boundary conditions. *Duderstadt et al.* [2014; 2016a] used incomplete, poorly-constrained and incorrect initial and boundary conditions, and they continue to advocate on the basis of uncertain results. Meanwhile, *Smart et al.* [2014] identified a series of ice core nitrate spikes that have since been confirmed in $^{10}$Be by *McCracken and Beer* [2015]. *Melott et al.* [2016] computationally reproduced the ionization profile of the only major balloon measurements to date. We show that our calculated nitrate enhancement is consistent with measured results, given current levels of uncertainty, and that extreme SPEs can potentially produce occasional nitrate spikes with hundreds of percent increases. Instead of repeating old arguments to dismiss nitrates as proxies of SPEs, it is past time for a dedicated, fine-resolution, multi-parameter, replicate ice core field campaign to resolve this debate.

# 1. Introduction

*Duderstadt et al.* [2016b] comment that, while listed as companion papers, *Melott et al.* [2016] and *Duderstadt et al.* [2016a] reached opposite conclusions and state that (1) they never saw *Melott et al.* [2016] before publication, and that (2) both we and *Sinnhuber* [2016] misrepresented their results. The first assertion is not strictly true since at least one of the co-authors reviewed our submitted paper. The second point is also debatable and we later address both issues in more detail. The *Melott et al.* [2016] finding that high-fluence, hard spectrum solar proton events (SPEs) may be detectable as short-duration, impulsive nitrate events (spikes) in ice cores is then dismissed by *Duderstadt et al.* [2016b] alleging that our results did not consider the $HNO_3$ background reservoir in their model. This too is incorrect. In fact, *Melott et al.* contended that both *Duderstadt et al.* [2014 and 2016a] studies were flawed, in part, because they overestimated this same reservoir. Our reply is structured into the same 4 sections as in *Duderstadt et al.* [2016b]. We defend the Melott et al. [2016] methodology in section 2, while questioning theirs, address alleged misrepresentations of their work in section 3, and point out a number of misrepresentations of our work in all 4 sections of *Duderstadt et al.* [2016b].

# 2. Why pre-existing stratospheric nitric acid was not neglected

*Duderstadt et al.* [2016b] assert that *Melott et al.* [2016] did not consider the "substantial pre-existing column of nitric acid [$HNO_3$]" and thus "incorrectly approach" the issue of detecting nitrate spikes from SPEs in polar snow. In so doing, they do not address *Melott et al.*'s previous arguments that the *Duderstadt et al.* [2014; 2016a] studies were flawed, because they overestimate their model column with modern stratospheric $NO_y$ levels and the percent enhancement approach cannot fully represent ground truth measurements with present uncertainties.

*Duderstadt et al.* [2016b] present Figures 1 and 2 to assert that lower stratosphere, background column densities of $NO_y$, mostly $HNO_3$, in their modeled results for 15 December 2004 and 15 February 2005 are roughly 20x larger than that produced by the 23 Feb 1956 SPE and could only result in at most a 5% (factor of 1.05) enhancement of nitrate above background in the snow, rather than a roughly 100% (factor of 2) enhancement, which they assert we measured, and certainly not the 400% (factor of 5) enhancement of some nitrate spikes. Their approach is incorrect for several reasons. First, *Duderstadt et al.* [2016b] misrepresent and inflate our results, by comparing the apex of the GISP2H ice core nitrate spike for winter 1956 with the local background. The correct method, using the spike integrated (or averaged) over its 4 data points yields a factor of 1.5 or a 50% enhancement above background, thus reducing the implied discrepancy from a factor of 20-80 to a factor of 10 (i.e. 5% allegedly expected, versus 50% measured).

Second, they compare modern levels of stratospheric $HNO_3$ with the $HNO_3$ expected from the 23 February 1956 SPE that was the subject of our study. *Melott et al.* [2016] stated: "As the WACCM model adopted by (*Duderstadt et al.*, [2016a]) is set to 2004 [*Brakebusch et al,* 2013; *Lamarque et al.,* 2012], it uses a much larger $NO_y$ background or reservoir than actually existed during the early and prehistoric SPEs that are of interest." Modern satellite data reflect the huge nitrate burden from fertilizer, the chemical industry and biomass burning and are inappropriate for disputing a potential SPE proxy for the pre-industrial era because they measure far more anthropogenic $NO_y$ and $HNO_3$ in the stratosphere [*Jacob*, 1999] than were present in the 1940s and 1950s. The 2004 WACCM run shown in their Figure 1 clearly includes stratospheric constituents polluted by anthropogenic sources; leaving out the bottom 8 km does not address this issue. The *doubling to tripling* of nitrate levels in Summit, Greenland snow since the 1950s, as noted by *Hastings et al.* [2009], was almost entirely caused by increases in anthropogenic-produced $HNO_3$ [Geng et al., 2014] and a roughly ~15% rise in $N_2O$ levels since the early 1950s further reduces the alleged discrepancy from a factor of 20-80 to a factor of 9. In addition, the large size of the modern $NO_y/HNO_3$ reservoir compared to winter $NO_3$ deposition levels and the high rate of anthropogenic inputs compared to natural sources suggests that current levels in the lower stratosphere may have increased from those in 1850 - 1950 by an even larger factor.

Although there is considerable uncertainty in these estimates, and the uncertainty range can be larger than the estimate (e.g. [*Jacob*, 1999], Table 10.1), the discrepancy is clearly much smaller than alleged by *Duderstadt et al.* and, given current unknowns and questions with this approach, may not exist if additional factors or mechanisms are involved, as we discuss below. As a final point on this subject, we note that $HNO_3$ levels are much lower in the Antarctic stratosphere, where it is colder and anthropogenic sources have had a negligible effect. Hence, an inland location with moderately high snow accumulation in Antarctica should offer the best place to obtain SPE statistics from ice core nitrate levels.

*Duderstadt et al.* [2016b] also misrepresent our 1956 impulsive nitrate event deposition period as "1-2 months" or "even shorter", citing *Smart et al.* [2014], thereby suggesting that this is too short a time to move all of the SPE-derived atmospheric $HNO_3$ into surface snow. *Smart et al.* [2014] assumed linearity between time and yearly deposition and actually estimated about 1 to 1.1 months (30 to 34 days), *not shorter*, while *Melott et al.* [2016] assumed lower winter snowfall [*Dibb and Fahnestock*, 2004] and repeatedly estimated 2-4 months, *not 1-2 months*. It is important to point out that *Melott et al.* [2016] was a proof of concept or feasibility study showing the deficiencies in previous modeling efforts and demonstrating for the first time that sufficient additional nitrate was produced by an SPE to explain the winter 1956 nitrate spike observed in the GISP2H ice core from Summit, Greenland.  We also noted that $HNO_3$ deposition processes involving polar stratospheric clouds (PSCs) and their response to SPEs are still poorly understood [*Engel et al.*, 2013; *Mironova et al.*, 2008; *Popp et al.*, 2006; *Yu*, 2004].  Accordingly, we did not attempt to prove that the $HNO_3$ produced by the February 1956 SPE between the surface and 46 km would all be incorporated into the ice sheet, or to propose a specific mechanism for this, as doing so was beyond the scope of our paper.  Absence of a known mechanism does not constitute proof of the null hypothesis; there were no known mechanisms initially to explain the very real springtime ozone hole discovered over Antarctica.  *Mironova et al.* [2008] found anomalous variations in sulfate and nitrate aerosol optical depths by several orders of magnitude in the Antarctic region upper troposphere 2 days after the 20 January 2005 SPE and noted that fundamental mechanisms to explain this phenomenon also are still poorly understood.  Potential mechanisms and specific details obviously require further research, including new ice core studies.  We reiterate that our findings appear plausible, given current uncertainties.  While a mechanism was not required for our limited study, one possibility was offered by *Yu* [2004] who postulated that cosmic rays might induce the freezing of supercooled droplets containing $HNO_3$ and explain the anomalous formation of large nitic acid trihydrate (NAT) particles at temperatures above the frost point that have been observed over wide Arctic regions [*Fahey et al.*, 2001].  This enhanced large NAT particle formation might apply specifically to the excess $HNO_3$ deposition in winter 1956.  NAT particles likely caused a 50% depletion of the Arctic $NO_y$ reservoir between 16 and 22 km in February 1995 Arctic [*Waibel et al.*, 1999].

### 3. Addressing alleged misrepresentations by *Melott et al.* [2016] and *Sinnhuber* [2016]

*Duderstadt et al.* [2016b] complain of listing *Duderstadt et al.* [2016a] and *Melott et al.* [2016] as companion papers, while not having had the opportunity to see *Melott et al.* [2016] prior to publication, and that both *Melott et al.* [2016] and *Sinnhuber* [2016] have made "significant misrepresentations" of their work. Regardless of whether misrepresentations have occurred, there were opportunities to address them during the publication process that were missed. *Duderstadt et al.* [2016a] and *Melott et al.* [2016] were initially submitted around the same time and were subsequently vetted by the same reviewer (Miriam Sinnhuber). It appears that the editor decided to publish the two papers together sometime after a second round of reviews of the *Melott et al.* [2016] paper. With a third reviewer's report in this round not returned, we were asked to respond to the first two reviewers' reports, both of which recommended publication with minor modifications. Shortly before the deadline for submitting our revised paper and reviewer responses, we were advised that the missing review had been received. Due to the controversial nature of this subject, the reviewer waived anonymity, revealing himself as Stanley Solomon (co-author of *Duderstadt et al.* [2014; 2016a], and announced the newly accepted (in press) *Duderstadt et al.* [2016a] paper, on which he was also co-author. The reviewer no longer recommended against publication of *Melott et al.*, as before, but rather "return for major revision" and he demanded that we address additional issues raised in the new (not yet published) work by *Duderstadt et al.* [2016a]. Hence, the editor asked us to respond accordingly and extended our deadline. We include this information merely to note that, if the review had been returned before *Duderstadt et al.* [2016a] was accepted for publication, there would have been an opportunity to respond to our paper and point out any misrepresentations.

Regarding the alleged misrepresentations, *Duderstadt et al.* [2016b] reiterate that *Duderstadt et al.* [2016a] defined $NO_y$ as "$N+NO_2+NO_3+2N_2O_5+HNO_3+HO_2NO_2+ClONO_2+BrONO_2$" and add "The WACCM chemical mechanism used in *Duderstadt et al.* [2016a] includes only these listed $NO_y$ species and reactions and does not include more extensive organic tropospheric chemistry involving PAN and alkyl nitrates..." with total $NO_y$ as alleged by *Melott et al.* [2016]. While we take them at their word, their documentation was incomplete and this could not be determined previously by a careful review of their paper. In fact, there was evidence suggesting otherwise and questions remain. First, their definition was confusing. Typically, $NO_y$ does include the $NO_x$ subfamily ($NO_x = NO+NO_2$). NO does not appear anywhere in their definition but we assume that it must have been used in their model. Second, many standard definitions of $NO_y$ do include PAN and organic nitrates (e.g. *Jones et al.* [1999], *Fahey et al.* [1985] and *Penkett et al.* [2009]), but apparently they were neglected completely in the WACCM model. Third, it is not clear exactly which version of WACCM was used; the most recent version of CESM (version 1.2.2, of which WACCM is a component) at the website cited in *Duderstadt et al.* [2014] ([https://www2.cesm.ucar.edu](https://www2.cesm.ucar.edu)) appears to include C-containing N (e.g. PAN) compounds in the $NO_y$ group, and this was the basis for our conclusion and our first objection to their percent enhancement approach, on which they focus. Regardless of whether PAN and other organic nitrates were involved in their model studies our second objection and the bottom line, which they ignore, is that the modern stratospheric $NO_y$ reservoir includes anthropogenic sources and is much higher than in pre-industrial times.

*Duderstadt et al.* [2016b] also allege a significant misrepresentation by both *Sinnhuber* [2016] and *Melott et al* .[2016] of their low ionization rates below 20 km. In doing so, *Duderstadt et al.* [2016b] misrepresent *Melott et al.* [2016] by quoting us out of context as saying that there was "little to no ionization below 20 km" in their calculations.    The complete statement, which referred to results from the balloon measurements of the 11 April 2013 SPE by *Nicoll and Harrison* [2014] showing an ionization peak around 20 km, was "None of the events modeled by (*Duderstadt et al.* [2016a]) show these characteristics, indicating, instead, peak ionization near 50 km or above for most events they studied, and little to no ionization below 20 km, with the notable exceptions of their 'hypothetical' hard and soft events based on the February 1956 and August 1972 SPEs."  While we are not obligated to defend statements by Sinnhuber, it is instructive to examine the source of this issue and why it appears in two separate publications. *Sinnhuber* [2016] stated "ionization rates in the lowermost stratosphere and troposphere (below ~20 km) are higher in the *Melott et al.* [2016] scenario for 1956 than in any of the scenarios shown in *Duderstadt et al.* ... possibly because the events studied by (*Duderstadt et al.*, [2016a]) have a less hard spectrum than the 1956 event." Ionization right down to the bottom of the troposphere can be significant [Thomas, B. C., E. E. Engler, M. Kachelrieß, A. L. Melott, A. C. Overholt, and D.V. Semikoz (2016) Terrestrial Effects of a Nearby Supernova in the Early Pleistocene, *Astrophys. J. Lett.,* submitted, arXiv:1605.04926] and must be done correctly. Figure 12 in *Duderstadt et al.* [2016a] presents ionization profiles of 6 SPEs, 4 "real" and 2 "hypothetical."  None of the real SPE events to which *Melott et al.* and *Sinnhuber* were referring show significant ionization below 20 km and 2 events appear to be truncated at or above 20 km. Figure 2 in *Duderstadt et al.* [2016a], which is difficult to interpret, also shows SPEs with 0-100 cm$^{-3}$ s$^{-1}$ ionization rates below 20 km most of the time. "Adjusting contours levels" and replotting as Figure 3 [*Duderstadt et al*., 2016b] suggests that the details were not clearly presented originally.

Figure 5 in *Melott et al.* [2016] was originally derived using a rough translation from ion pairs $(cm^{-2} s^{-1})$ to $(cm^{-3} s^{-1})$ for comparison purposes and after publication was discovered to plot numbers ~3x too high. While relative differences between the data, and more importantly our final deposition numbers were not affected, the values have been recalculated using the correct methodology and are presented here in Figure 1. Even when the corrected values for 23 Feb 1956 are presented, ionization rates below 20 km (~2 - 4000 $cm^{-1} s^{-1}$) are higher than in any of *Duderstadt et al.*'s [2016a] modeled "real" SPEs results (0 - ~3000 $cm^{-1} s^{-1}$).

*Duderstadt et al.* [2016b] then present Figure 3 showing ionization rates for the 20 January 2005 SPE to support their assertion of misrepresentation and claim that their modeled ionization levels below 20 km are consistent with *Nicoll and Harrison* [2014]. However, Nicoll and Harrison measured the 11 April 2013 SPE where values peaked near 20 km, while *Duderstadt et al.*'s [2016b] Figure 3 peaks near 40 - 50 km (as originally shown in *Duderstadt et al.* [2016a], Figure 2). Comparing a simulation of one event with measurements from another event that produce a maximum at a completely different altitude does not show the methodology is correct. The only way to prove that their model replicates the results of *Nicoll and Harrison* [2014] is to run it for the exact event. Hence, the complete statements on ionization rates by both *Melott et al*. and *Sinnhuber*, appear to be substantially correct.

## 4. Conclusions

*Duderstadt et al.* [2016b] advocate cosmogenic isotopes in ice cores for SPE proxies as an alternative to nitrates, citing a promising study of $^{10}Be$ by *McCracken and Beer* [2015] who independently identified *3 of the same 4 separable ground level events* (or 4 of 5 total; 2 in 1942 were too closely spaced to resolve separately) between 1940-1956 that were previously identified as nitrate spikes by *Smart et al.* [2014] in finely-resolved ice cores. This is strong confirmation of the fine-resolution nitrate approach, and it is clear that the two different methods have complementary value. As we have emphasized before, modeling and comparison can constrain SPE fluence and spectral information.

*Duderstadt et al.*'s [2016b] assertion that the many other causes of nitrate variability will not allow proxy records of SPEs to be distinguished is simply incorrect.   *Wolff et al.* [2016] have shown that collocated, multi-species measurements that include full ionic balance, nitrate concentration, and liquid conductivity can be used to separate nitrate spikes in the ice sheet formed by high-conductivity, SPE-produced $HNO_3$ from those formed by low-conductivity aerosol sources (biomass burning, sea salt and dust). As previously pointed out in *Smart et al.* [2014], *Wolff et al.* [2008] also noted that in winter sea salts are unlikely to be the cause of nitrate spikes at the remote Summit location.  This also applies to biomass burning.  High conductivity levels, no apparent biomass burning indicators, and time of year were the bases for our finding that the winter 1956 nitrate spike in Summit ice was composed of $HNO_3$ and likely produced by the February 23 1956 SPE.  The real problem is that no multi parameter measurements have been undertaken on replicate ice cores at the fine resolutions required to determine conclusively nitrate spike sources.  We have repeatedly called for such a study (*Smart et al.* [2014], *Smart et al.* [2016], *Melott et al.* [2016]), yet some members of the ice core community continue to resist this strongly using circular reasoning (by concluding in advance) that it will yield no useful information and is therefore unnecessary.

*Melott et al.* [2016] moved this debate forward by clearly showing past assumptions involving major impulsive nitrate events and their association with SPEs needed revision and updating.

*Duderstadt et al.* [2016b] comment on issues that are no longer relevant stating "Traditional arguments for associating SPEs with nitrate in ice cores selectively chose spikes that fall near dates of observed, historical flares... These selected spikes are then used to extrapolate other nitrate spikes to hypothetical storms."  We have previously acknowledged that Table 1 of *McCracken et al.* [2001], to which their statement refers, needs revision both with respect to the calibration of nitrate spikes to higher energy particles and to culling out those spikes with low levels of conductivity.  A first pass at this has been completed recently and almost half of the SPEs remain [*Smart and Shea*, personal communication, 2016].

*Duderstadt et al.* [2016b] highlight the largest nitrate "peaks" in the Summit GISP2 and BU ice core records ("typically hundreds of percent increases") that "Even extreme SPE enhancements cannot explain" for which *Melott et al.* [2016] made no claims.  They do not distinguish between true nitrate "spikes", which few studies have resolved to date, and the well-known annual nitrate cycle in the GISP2H and BU ice cores, which can produce summer "peaks" that vary ~50% - 400% above the winter background (i.e. *Dreschhoff and Zeller* [1994 and 1998] and *Kepko et al.* [2009]), especially since 1950 when anthropogenic sources and inputs became significant.  We do not allege that all, or even most, impulsive nitrate events recorded in ice cores, including those with hundreds of percent increases, are attributable to SPEs.  We do suggest that if the 23 February 1956 SPE could produce a nitrate spike with a 50% increase above local background, an equivalent, hard-spectrum SPE with a 10x larger fluence, as may have occurred in 775 AD and 994 AD [*Melott and Thomas,* 2012; *Usoskin et al.*, 2013; *Miyake et al.*, 2013; *Mekhaldi et al.*, 2015; *Ding et al.*, 2015], could potentially produce a nitrate spike in ice cores with a 500% increase that would be easily visible (i.e. *Melott et al.*, [2016], Figure 7).  It is SPEs of the scale of 1956 and larger that pose the greatest threat to our technological civilization and are of the greatest interest for deriving SPE statistics.

*Duderstadt et al.* [2016b] complain that "traditional" arguments selectively chose spikes that fall near dates of observed, historical flares in the 1940s and 1950s and that these selected spikes are then used to extrapolate other nitrate spikes to hypothetical storms. While this was done years ago by *McCracken et al.* [2001] and a few others more recently, it has nothing to do with *Melott et al.* [2016]. *Duderstadt et al.* [2016b] also state: "The authors study only the nitrate contributions that support their conclusions while neglecting all other material in the surrounding medium" and ignore "alternative explanations." Again this is incorrect; we merely came to different conclusions. We specifically considered organic and anthropogenic sources of nitrate and concluded the SPE signature is expected to be from gaseous $HNO_3$ that is subject to relatively rapid deposition (~ one week in the troposphere). Comparing the SPE-produced $HNO_3$ to modern levels dominated by anthropogenic sources that were minor in 1956 and not measurable before 1850 [*Geng et al.*, 2014] is inappropriate. We also considered sources posited by Wolff et al. [2012; 2016], noting that their coarse resolution ice core studies did not resolve any nitrate spikes, thus pointing to the need for fine-resolution studies [*Smart et al.*, 2014]. Merely because broad summer nitrate "peaks" can be "associated" with and "explained" by other tropospheric sources that mostly peak in summer, this does not prove that they are the cause of all or even most nitrate spikes. *Duderstadt et al.* appear to want it both ways. On the one hand they assert that all nitrate spikes result from biomass burning and other tropospheric sources. At the same time, they note that $HNO_3$ is the overwhelming component of $NO_y$ in the lower stratosphere in Arctic winter. This implies that $HNO_3$ may be the major source of winter nitrate deposition at Summit, not biomass burning, sea salt or dust.

In closing, *Duderstadt et al.* [2016b] have not shown convincingly that our methodology is fundamentally flawed. As noted by *Sinnhuber* [2016], it is not clear that assuming a given percentage increase in nitrate levels in the snow must be matched by at least an equivalent increase in stratospheric $HNO_3$ to demonstrate SPE causation is the more realistic approach. It is clear that this controversy has not been resolved. Using incomplete understanding, poorly constrained models and old arguments to dismiss nitrates in ice cores as proxies of SPEs does not move the debate forward. A thorough analysis of nitrate spike sources and possible SPE signatures will require a dedicated field campaign.

**Acknowledgments**

We are grateful for helpful comments from M. Shea and D. Smart. I. Usoskin [personal communication 2015] provided SPE-ionization data. ALM and BCT were supported by NASA grant NNX14AK22G.

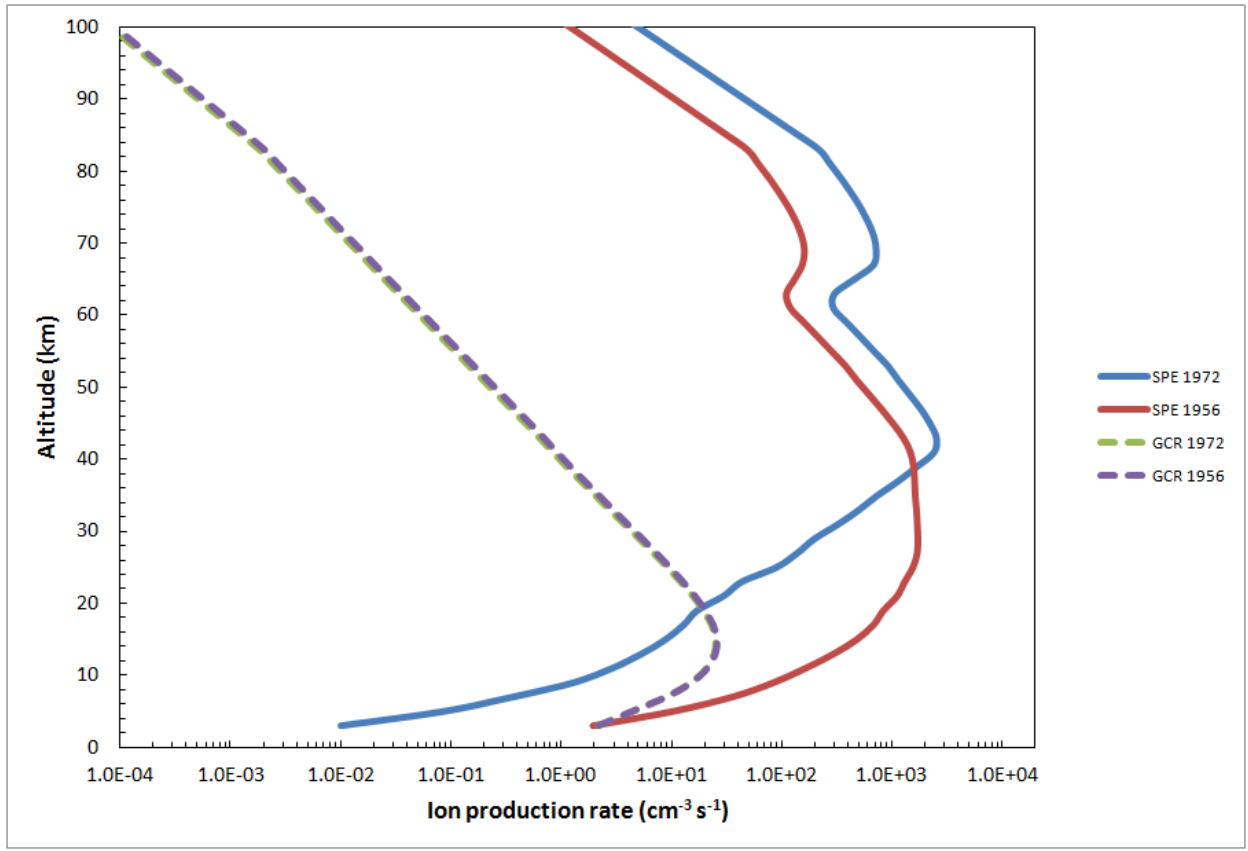

**Figure 1.** Figure 5 from *Melott et al.* [2016] revised. Ionization rate profiles (provided by I. Usoskin). The dashed lines are the background ionization rates (nearly identical) in the atmosphere due to Galactic Cosmic Rays in February 1956 and August 1972. The solid lines are ionization rates assuming a 1-day event duration. The red solid line shows ionization due to the 23 February 1956 SPE. Note that there is considerable ionization below 20 km, which has not been shown or included in many computations. The blue solid line shows ionization due to the 3-4 August 1972 SPE.